\documentclass[pdflatex,sn-mathphys-num]{sn-jnl}

\DeclareUnicodeCharacter{202F}{\textbf{FIX ME!!!!}}

\usepackage{graphicx}%
\usepackage{multirow}%
\usepackage{amsmath,amssymb,amsfonts}%
\usepackage{amsthm}%
\usepackage{mathrsfs}%
\usepackage[title]{appendix}%
\usepackage{xcolor}%
\usepackage{textcomp}%
\usepackage{manyfoot}%
\usepackage{booktabs}%
\usepackage{algorithm}%
\usepackage{enumitem}
\usepackage{algorithmicx}%
\usepackage{algpseudocode}%
\usepackage{listings}%
\usepackage{geometry}
\usepackage{array}
\theoremstyle{thmstyleone}%
\theoremstyle{thmstyletwo}%

\theoremstyle{thmstylethree}%

\begin{document}

\title[Article Title]{ Post Quantum Cryptography \& its Comparison with Classical Cryptography }


\author{\fnm{Tanmay Tripathi}}
\author{\fnm{Abhinav} \sur{Awasthi}}
\author{\fnm{Shaurya} \sur{Pratap Singh}}
\author{\fnm{Atul} \sur{Chaturvedi}}
\affil{\orgname{Pranveer Singh Institute of Technology, Kanpur, India}}


\abstract{Cryptography plays a pivotal role in safeguarding sensitive information and facilitating secure communication. Classical cryptography relies on mathematical computations, whereas quantum cryptography operates on the principles of quantum mechanics, offering a new frontier in secure communication. Quantum cryptographic systems introduce novel dimensions to security, capable of detecting and thwarting eavesdropping attempts. By contrasting quantum cryptography with its classical counterpart, it becomes evident how quantum mechanics revolutionizes the landscape of secure communication. }




\maketitle

\section{Introduction}

Classical cryptography, a time-honored method for safeguarding communication and data, leans on mathematical algorithms and computational intricacy to encode and decode messages. Its roots extend over centuries, with historical examples like the substitution ciphers such as the Caesar cipher and transposition ciphers like the Rail Fence cipher. These classical cryptographic techniques typically employ keys to encrypt and decrypt messages, with the security often contingent upon the confidentiality of the key. Notable classical cryptographic algorithms include the RSA algorithm for public-key encryption, the AES algorithm for symmetric-key encryption, and the Diffie-Hellman key exchange protocol. Nevertheless, classical cryptography confronts potential threats posed by advancements in computing power and the evolution of new mathematical techniques, particularly with the rise of quantum computers. Quantum computers possess the capability to compromise many existing cryptographic schemes, such as RSA and ECC, leveraging their efficiency in solving specific mathematical problems like integer factorization and discrete logarithms. Post-Quantum Cryptography: Post-Quantum Cryptography (PQC) represents a contemporary cryptographic approach crafted to withstand assaults from both classical and quantum computers. Its objective is to forge cryptographic algorithms resilient even in the face of potent quantum computers. PQC algorithms typically derive from mathematical problems deemed arduous for both classical and quantum computers to solve. Examples are lattice-based cryptography, code-based cryptography, hash-based cryptography, and multivariate polynomial cryptography \cite{b1}. The evolution of PQC is paramount for preserving the enduring security of sensitive information in an era anticipated to witness the emergence of quantum computers. Endeavors to standardize PQC algorithms are underway, aiming to ensure broad adoption and compatibility across diverse systems and applications. As interdisciplinary collaborations progress, the goal is to establish new cryptographic standards that ensure the enduring security of information in the post-quantum era \cite{b2, b3, b4, b5, b6, b7, b8, b9, b10, b11, b12, b13}.\\
Classical cryptography is grounded in mathematical algorithms and computational intricacies to ensure secure communication. In contrast, post-quantum cryptography strives to address the potential weaknesses of classical cryptographic algorithms and withstand attacks from both classical and quantum computers \cite{b14, b15, b16}.

\subsection{Historic Timeline}

In the early 1970s, IBM established a 'crypto group' that developed a block cipher to safeguard its clients' data. In 1973, the United States adopted it as a national standard, known as the Data Encryption Standard (DES), which remained in use until its vulnerability was exposed in 1997.  In 1976, Whitfield Diffie and Martin Hellman introduced the concept of the Diffie-Hellman key exchange, revolutionizing cryptography by dynamically generating a pair of keys (one public, one private but mathematically linked) for each correspondence, eliminating the need for pre-arranged code keys. Quantum cryptography was initially proposed in 1984 by Bennett and Brassard. In 2000, the Advanced Encryption Standard (AES) superseded DES, providing enhanced security. AES employs symmetric-key encryption, requiring both the user and sender to possess the same secret key. Public Key Infrastructure (PKI) is a broad term encompassing solutions for creating and managing public-key encryption. In 2005, Elliptic Curve Cryptography (ECC) emerged as an advanced public-key cryptography scheme, enabling shorter encryption keys. ECC offers heightened security compared to RSA and Diffie-Hellman, as elliptic curve cryptosystems are more resistant to attacks. 

\subsection{Mathematics in Cryptography}

\textbf{One way function: }A one-way function in cryptography serves as a fundamental tool, offering a mathematical process that is straightforward to execute in one direction but exceedingly challenging to reverse. Put simply, when provided with an input, it's effortless to calculate the corresponding output. However, when presented with the output alone, it becomes computationally impractical, if not impossible, to discern the original input without access to particular knowledge or supplementary data. \\

Characteristics of One Way Function:
\begin{itemize}
	\item Easy to Compute
	\item Hard to Invert
\end{itemize}

\textit{One Way functions are building blocks in many cryptographic protocols.}

\textbf{Hash Functions: }One-way hash functions play a pivotal role in cryptography, serving various purposes such as securely storing passwords, generating message digests for data integrity verification, and deriving cryptographic keys. Examples of commonly used hash functions include SHA-256 and SHA3. 

\textbf{Trapdoor Functions: }Certain one-way functions possess a unique property known as a trapdoor, enabling efficient computation of the inverse under specific conditions. These trapdoor functions are integral to asymmetric encryption schemes like RSA and Diffie-Hellman key exchange, facilitating secure communication and key establishment over insecure channels.

\textbf{Digital Signatures: }In digital signature algorithms, one-way functions are utilized to generate and verify signatures, thereby ensuring the authenticity and integrity of digital documents. By applying oneway functions to the message along with a private key, a digital signature is generated, which can be verified using the corresponding public key, providing assurance of the document's origin and integrity. 

\textbf{Use of One Way Function}
\begin{enumerate}
	\item \textbf{Password Hashing: }One-way hash functions are extensively employed for securely storing passwords in databases. When a user creates or updates their password, it undergoes hashing using a one-way function before being stored. During authentication, the input password is hashed again, and the resulting hash is compared to the stored hash. Even if the database is compromised, retrieving the original passwords from the hashed values is computationally challenging. 
	
	\item \textbf{Message Integrity: } Verification: One-way hash functions are utilized to generate message digests, which are then utilized for data integrity verification. By hashing a message and comparing its digest to a previously computed value, recipients can ascertain that the message hasn't been tampered with during transmission. This technique finds common use in protocols like TLS/SSL and digital signatures. 
	
	\item \textbf{Cryptographic Key: }Derivation: One-way functions are pivotal in deriving cryptographic keys from secret values or passwords. Key derivation functions (KDFs) leverage one-way functions to derive secure cryptographic keys from initial inputs, ensuring that these keys are computationally indistinguishable from random values.
\end{enumerate}

\section{Classical Cryptography}
Cryptography, derived from the Greek words "crypto" meaning secret and "graphy" meaning writing,      encompasses the study of secure communication techniques. In classical cryptography, encryption relies primarily on the complexity of mathematical operations. Specifically, the security of classical encryption methods hinges on the computational challenge of factoring large. Classical Cryptography has two types of techniques:
\begin{itemize}
	\item \textbf{Symmetric Cryptography: }In symmetric cryptography, a single key serves both for encrypting and decrypting data.
	
	\item \textbf{Asymmetric Cryptography: }In asymmetric cryptography, a pair of keys, namely the public key and the private key, are employed for encryption and decryption purposes. 
\end{itemize}
Throughout the paper, the sender is denoted as 'Alice,' the recipient as 'Bob,' and any potential eavesdropper as 'Eve.' The most popular among them that are adopted globally have been described below \cite{b14, b15}.

\textbf{Substitution Ciphers}\\
Substitution ciphers are cryptographic techniques that entail substituting plaintext characters with different ciphertext characters based on a predetermined system. The Caesar cipher, a notable example, shifts each letter of the alphabet by a fixed number of positions. Additionally, other substitution ciphers such as the Atbash cipher and the Polybius square offer alternative methods for encoding plaintext into ciphertext.
~\\
\textbf{Data Encryption Standard (DES)}\\
DES, also known as Data Encryption Algorithm (DEA) in the United States or Data Encryption Algorithm-1 (DEA1), is a symmetric-key encryption algorithm employed for securing electronic data. Originally developed by IBM in the 1970s and subsequently endorsed by the U.S. government as a federal standard for encryption in 1977, DES aimed to facilitate secure communication over insecure channels. The encrypting and decrypting algorithms of DES are publicly disclosed. In DES, data encryption occurs using a secret key, which is also utilized for decryption. However, DES has become susceptible to cryptographic attacks due to its relatively short 56- bit key length, rendering it vulnerable in modern cryptographic contexts. Consequently, DES has largely been supplanted by more robust encryption algorithms such as the Advanced Encryption Standard (AES).
~\\
\textbf{Public Key Cryptographic (PKC) Systems}\\
A public key cryptosystem is a cryptographic technique that employs two keys: a public key and a private key. These keys are mathematically linked so that data encrypted with one key can only be decrypted with the other [3].
Here's how it operates:
\begin{itemize}
	\item  \textbf{Public Key: }This key is openly available and is used for encryption. Any sender wanting to encrypt a message for the owner of the public key can do so using this key.
	\item \textbf{Private Key: }This key is kept confidential by its owner and is employed for decryption. Only the owner of the private key can decrypt messages encrypted with their public key.
\end{itemize}

The security of the system relies on the computational complexity of specific mathematical problems, such as factoring large numbers or computing discrete logarithms. Public key cryptosystems \cite{b19} offer numerous advantages, including secure communication over insecure channels, digital signatures, and key exchange protocols. One of the most prominent public key cryptosystems is Rivest - Shamir - Adleman (RSA), developed in the 1970s and still widely utilized today. Other examples include Diffie-Hellman key exchange and elliptic curve cryptography.\\
For example, a software publisher can create a signature key pair and include the public key in software installed on computers. Later, the publisher can distribute an update to the software signed using the private key, and any computer receiving an update can confirm it is genuine by verifying the signature using the public key. As long as the software publisher safeguards the private key, even if a malicious actor manages to distribute harmful updates to computers, they cannot persuade the computers that any of these updates are legitimate \cite{b20}.
~\\
\textbf{One-time pad (OTP) Cryptosystem}\\
The one-time pad cryptosystem \cite{b21} created by Gilbert Vernamin 1917 is very simple and yet, very effective. The system ensures perfect secrecy. A one-time pad is sometimes considered as ideal cipher. The system is named after encryption method in which a large, non-repeating set of keys is written on sheets of paper, attached together into a pad. One-time pads \cite{b20} have been used when both parties started out at the same physical location and then separated, each with knowledge of the keys in the one-time pad. The key utilized in a one-time pad is referred to as a secret key because if it is disclosed, the messages encrypted with it can be easily deciphered.

\section{Post Quantum Cryptography}

Post-quantum cryptography (PQC) is a specialized field within cryptography dedicated to crafting cryptographic systems that maintain their security even when faced with the formidable computational power of quantum computers. Quantum computers possess the capability to potentially 
dismantle several prevalent cryptographic algorithms, including RSA and ECC, leveraging their adeptness at swiftly solving specific mathematical challenges like integer factorization and discrete logarithms.
Post-quantum cryptography (PQC) endeavors to furnish fundamental cryptographic tools,encompassing encryption schemes, digital signatures, and key exchange protocols, engineered to withstand assaults originating from both classical and quantum computers. 
Post-quantum cryptography (PQC) is imperative to safeguarding sensitive data, such as financial transactions, communication channels, and data privacy, ensuring their long-term security as quantum computers emerge into reality. Additionally, quantum cryptography protocols often employ quantum key distribution (QKD) schemes, such as BB84 or E91,which enable two parties to generate a shared secret key with provable security guarantees. These keys can then be used for secure communication via classical encryption techniques.
~\\
\textbf{Quantum Key Distribution}\\
One of the most powerful and commonly used quantum protocols is the Quantum Key Distribution (QKD), which enables two parties to share a secret key over a quantum channel, such as a fiber-optic cable or free space. QKD is based on the principle of quantum entanglement.
~\\
\textbf{BB84 Protocol}\\
It is one of the earliest and most widely used QKD protocols, introduced by Charles Bennett and Giles Brassard in 1984. The protocol uses the properties of quantum mechanics, such as the polarization of photon, to establish a shared secret key between two parties.
~\\
\textbf{E91 Protocol}\\
E91 also called the Ekert Protocol, it is also a QKD Protocol, like BB84, that uses the properties of entangled particles to establish a shared secret key between two parties. It was introduced by the British-Polish professor of quantum physics at the Mathematical Institute, University of Oxford,
Artur Ekert in 1991.
~\\
In Quantum Key Distribution sequence of operations are as follows :-
Initially, Alice produces and transmits to Bob a series of photons,each characterized by randomly chosen polarizations (0, 45, 90, or 135 degrees). Upon reception, Bob randomly selects whether to measure rectilinear or diagonal polarization for each photon. Subsequently, Bob publicly discloses the type of measurement conducted (rectilinear or diagonal) for each photon, withholding the specific measurement outcome. Alice then openly informs Bob whether his chosen measurement type was correct for each photon. Subsequently, both Alice and Bob discard instances in which Bob's measurement was incorrect or in which his detectors failed to register a photon. measurement was incorrect or in which his detectors failed to register a photon.

\begin{figure}[htbp]
	\centerline{\includegraphics[scale=.2]{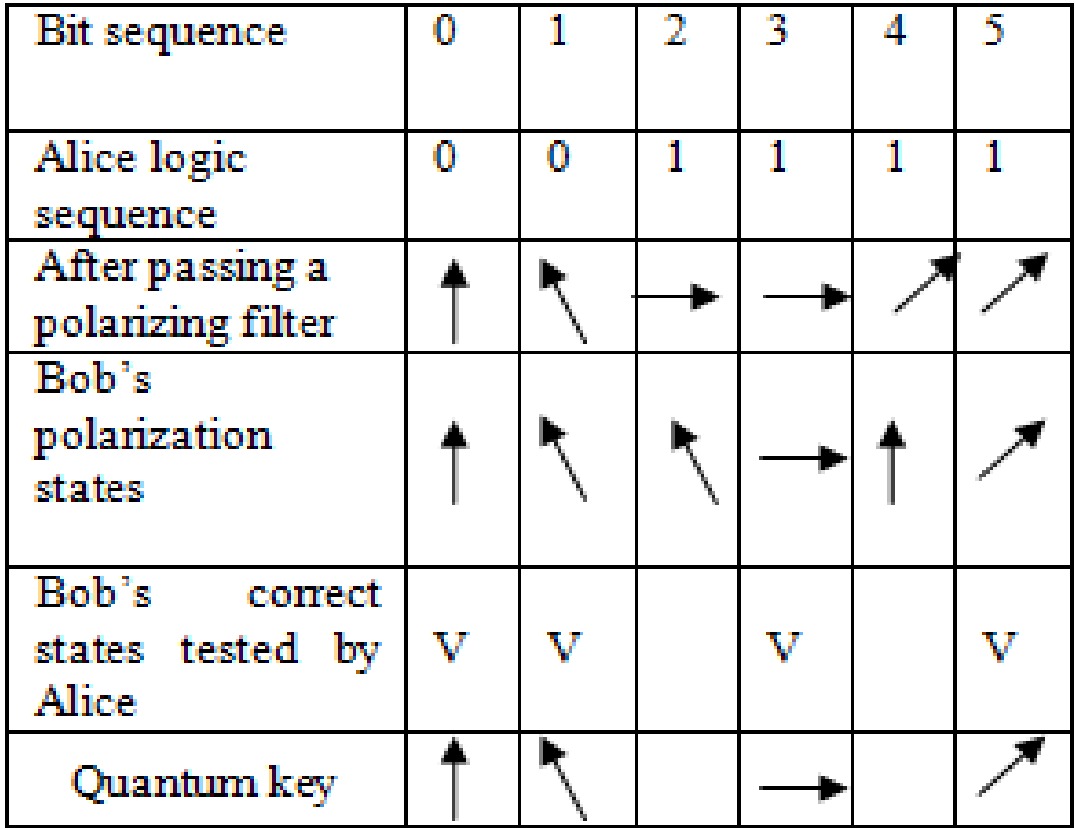}}
	\caption{Sequence of operation}
	\label{fig}
\end{figure}

If no eavesdropping has occurred on the quantum channel, the remaining polarizations become shared secret information between Alice and Bob. They proceed to test for eavesdropping by openly evaluating and discarding a randomly selected subset of their polarization data.. If this evaluation reveals evidence of eavesdropping, Alice and Bob discard all data and begin anew with fresh photons. Otherwise, they utilize the remaining polarizations as shared secret bits, interpreting 0-degree or 45- degree photons as binary 0s and 90-degree or 135-degree photons as binary 1s. In the event that Alice makes an incorrect measurement, she resends Bob a photon consistent with her measurement result. This action effectively randomizes the original polarization sent by Alice for that specific photon, resulting in errors in one-fourth of the bits in Bob's data subjected to attack. Since there is no information about Alice's secret choice, there is a $50\%$ chance (probability 1/2) of estimating the correct bit and a $50\%$ chance (probability 1/2) of estimating incorrectly. If the estimation is correct, Alice's transmitted bit is received with a probability of 1. However, if the estimation is incorrect, Alice's transmitted bit is received correctly with a probability of 1/2. Overall, the probability of accurately receiving Alice's transmitted bit is calculated as follows \cite{b20, b21}:
$$P=1.1/2+1/2.1/2=3/4$$
The BB84 scheme was adapted to create a functional quantum cryptography kit at IBM, with modifications aimed at addressing practical issues such as detector noise. Unlike the BB84 scheme, which encodes each bit using a single polarized photon, this kit encodes each bit in a faint flash of light\cite{b21}.
~\\
However, this modification introduces a new eavesdropping risk: if Eve intercepts the beam, she could split each flash into two weaker flashes, intercepting one while allowing the other to reach Bob. If Eve only diverts a small portion of the beam, Bob may not detect the weakened signal or may attribute it to expected losses in the channel. To mitigate this attack, the data transmission rate can be reduced by sending very dim flashes with an intensity of less than one photon per flash on average. Another challenge is that available detectors sometimes produce a response even when no photon has arrived, leading to errors even in the absence of eavesdropping. Additionally, a vulnerable aspect is key storage: once Alice and Bob establish the key, they must store it until needed. However, the longer the key remains stored, the more susceptible it becomes to unauthorized access. One potential solution is to develop a cryptosystem based on the well-known Einstein-Podolsky-Rosen (EPR) effect. Ekert recently devised a cryptosystem leveraging the EPR effect, which provides security for both key storage and key distribution. However, practical implementation is hindered by the technical challenge of storing photons for more than a fraction of a second.

~\\

\textbf{Two New Algorithm that are designed for Post Quantum Cryptography}

"Falcon" and "Crystal-Dilithium" are two distinct cryptographic algorithms proposed for post-quantum cryptography (PQC), a field dedicated to developing cryptographic methods resilient against quantum computer attacks.

\textbf{FALCON:} Falcon serves as a digital signature scheme grounded in the complexity of the Learning with Errors (LWE) problem. It's engineered to thwart attacks from both classical and quantum computers, aiming to furnish efficient and secure digital signatures applicable in real-world scenarios, even amidst the existence of quantum computers \cite{b18}.\\
\textbf{CRYSTALS-DILITHIUM: } Cryptographic Suite for Algebraic Lattices (CRYSTAL) encompasses a suite of cryptographic algorithms rooted in lattice problems. This suite comprises encryption schemes, key exchange protocols, digital signature schemes, and hash functions. The objective behind CRYSTAL's design is to furnish cryptographic primitives impervious to quantum attacks, rendering it apt for deployment in a post-quantum cryptographic landscape.

Both Falcon and Crystal-Dilithium are integral components of the persistent endeavor within the cryptographic community to devise and standardize novel algorithms capable of withstanding assaults from quantum computers. Such advancements are crucial, considering the potential vulnerability of numerous existing cryptographic schemes to quantum-based attacks.[5]

\section{Classical V/S Post Quantum Cryptography}

\begin{table}[htbp]
	\caption{Post Quantum Cryptography V/S Classical Cryptography}
	\begin{tabular}{|c|c|c|}
		\hline
		\textbf{Features} & \textbf{Post Quantum Cryptography} & \textbf{Classical Cryptography}\\
		\hline
		Basis & Quantum mechanics & Mathematical computation\\
		\hline
		Development & Infantile \& not tested fully & Deployed and tested\\
		\hline
		Existing Infrastructure & Sophisticated & Widely used\\
		\hline
		Digital Signature & Not present & present\\
		\hline
		Bit rate & 1Mbit/s avg \cite{b22} & Depend on Computing power\\
		\hline
		Cost & Crypto chip $\texteuro100,000$ \cite{b23} & Almost zero\\
		\hline
		Register storage (n-bit) at any moment & One n-bit string & Million of miles\\
		\hline
		Communication Range & 10 miles max \cite{b21} & Million of miles\\
		\hline
		Requirements & Devoted h/w \& communication lines & S/w and portable\\
		\hline
		Life expectancy & No change based on physics laws & Require changes as computing power increases\\
		\hline
		Medium & Dependent & Independent\\
		\hline
	\end{tabular}
\end{table}

\section*{Known Attack on Classical and Post Quantum Cryptography}

\begin{itemize}
	\item \textbf{Attack on Classical Cryptography}
	
	\textbf{Known-Plaintext Attack: }This technique involves leveraging knowledge of plaintext-ciphertext pairs to deduce the encryption key. By analyzing the relationship between the known plaintext and its corresponding ciphertext, an attacker can potentially infer information about the encryption algorithm or the key itself.
	
	\textbf{Man-in-the-Middle Attack: }This attack occurs when a malicious actor intercepts communication between two parties, often without their knowledge. The attacker can then modify the communication or relay it to the intended recipient while capturing sensitive information exchanged between the parties. Man-in-the-middle attacks pose a significant threat to the confidentiality and integrity of communication channels \cite{b19}.
	
	\textbf{Birthday Attack: }The birthday attack exploits the probability of collisions in hash functions, particularly in cryptographic hash functions. By leveraging the birthday paradox, which states that in a group of 23 people, there is a greater than $50\%$ chance that two people share the same birthday, attackers can efficiently search for collisions in hash functions. This attack is particularly relevant in scenarios where hash functions are used for tasks such as digital signatures or data integrity verification \cite{b21}.
	
	\item \textbf{Attack on Post Quantum Cryptography}
	
	\textbf{Grover's Algorithm:} Grover's Algorithm is a quantum algorithm that addresses the problem of searching an unsorted database. It has the capability to perform a brute-force search with a complexity that is the square root of the number of items in the database. This algorithm has significant implications for symmetric key encryption as it can potentially reduce the time required to break cryptographic schemes relying on brute-force attacks \cite{b20}.
	
	\textbf{Shor's Algorithm:} Shor's Algorithm is another quantum algorithm with profound implications for cryptography. It is designed to efficiently factor large integers and solve the discrete logarithm problem. These tasks are notoriously hard for classical computers, forming the basis of many asymmetric encryption schemes such as RSA and ECC. Shor's Algorithm's ability to efficiently solve these problems poses a serious threat to the security of these cryptographic systems, potentially rendering them vulnerable to attacks by quantum computers \cite{b18}.
\end{itemize}

\section*{Conclusion}
In conclusion, classical cryptography has been effective and applicable in various contexts, yet the rise of quantum computing demands a shift towards post-quantum cryptographic methods. This transition is essential to fortify our digital infrastructure against potential threats posed by quantum computing advancements. Embracing post-quantum cryptography is pivotal for protecting sensitive data and upholding the reliability of communication channels in the foreseeable future.

\subsection*{Challenges \& Future Direction}
In the future, enhancing the performance of practical QKD systems and further improvements, both in key rate and secure transmission distance, are necessary for some applications. Another vital point is that, in real life, that is, quantum signals may share the channel with regular classical signals. The final goal is to achieve a client affable QKD system that can be effortlessly included in the Internet. To achieve a higher QKD key rate, one can consider other QKD protocols. Continuous variable QKD is projected to get a higher key rate in the small and medium transmission distance. Still, the scalability is a big challenge, as no one knows how to build a large scale quantum computer, which is interesting subject to be worked out.


\end{document}